\documentclass[twocolumn]{jpsj2}
\def\runauthor{T. \textsc{Yoshioka},
A. \textsc{Koga} and  N. \textsc{Kawakami} }

\title{
Frustrated Ising model on the garnet lattice
}

\author{%
Takuya Yoshioka, Akihisa Koga and Norio Kawakami
}

\inst{
Department of Applied Physics, Osaka University, Suita, 
Osaka 565-0871 
}

\recdate{\today}

\abst{
We investigate a frustrated Ising spin system on the garnet lattice
composed of a specific network of corner-sharing triangles.
By means of Monte Carlo simulations with the heat bath algorithm,
we discuss the magnetic properties at finite temperatures.
It is shown that the garnet spin system with the  
nearest-neighbor couplings does not exhibit any magnetic 
transitions, yielding the large residual entropy at zero temperature.
We also investigate the effect of the long-range dipolar interaction 
systematically to determine the phase diagram of the Ising model 
on the garnet lattice.  We find that there appear 
a variety of distinct phases
depending on the cutoff length of the long-range interaction,
 in contrast to the pyrochlore spin-ice systems.
} 

\kword{%
spin ice, frustration, garnet lattice, Ising model
}

\begin{document}
\sloppy
\maketitle

\section{Introduction}
Recently, geometrically frustrated spin systems have attracted 
much interest. One of the most remarkable examples is the
Ising-spin system on the geometrically frustrated
pyrochlore lattice, such as 
 $\rm Dy_2Ti_2O_7$ \cite{Dy2Ti2O7,Dy2Ti2O7_2,Siddharthan,Siddharthan2,
 spin_ice,Melko}
  and $\rm Ho_2Ti_2O_7
 $.\cite{Siddharthan,Siddharthan2,spin_ice,Melko,Ho2Ti2O7,Ho2Ti2O7_2}
In these compounds, a magnetic moment located at each 
lattice point is quantized, due to strong easy-axis anisotropy,
along the line joining the tetrahedra that compose the basic
unit of the pyrochlore lattice.
Strong geometrical frustration on this lattice yields 
remarkable properties
 quite analogous to the real ice,  now called the {\it spin ice}
 \cite{Ho2Ti2O7}:
the system does not show any magnetic phase transitions,
resulting in the finite residual entropy at zero temperature
\cite{spin_ice,Dy2Ti2O7}.
It has been further confirmed that some portion of 
the residual entropy  is released by an applied 
magnetic field, yielding another two-dimensional frustrated
system on the Kagome lattice, which is now referred to as the {\it Kagome ice}
\cite{kagome1,kagome2,kagome3}.
More recently, it has been argued that  the competition of
the long-range dipolar interaction and
the nearest neighbor superexchange interaction  plays a crucial 
role  to determine the magnetic properties in this class 
of frustrated Ising spin systems.\cite{Ho2Ti2O7_2,Melko}
These remarkable experimental observations of the 
spin-ice behavior have stimulated 
 further intensive investigations in this field.

\begin{figure}[htb]
\begin{center}
\includegraphics[width=7cm]{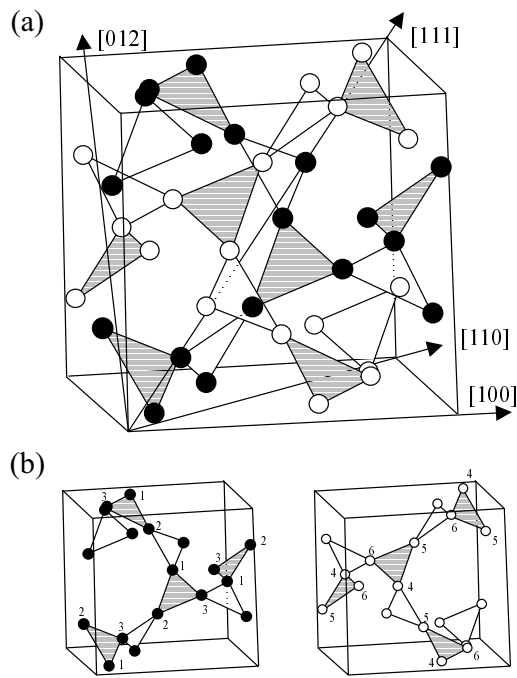}
\end{center}
\caption{(a) Garnet structure: there are 
twenty four sites in the unit cell.
$[100], [110], [111]$ and $[012]$ represent the directions of
 the applied magnetic field (see text).
(b) Two equivalent sublattices composing the garnet structure.
}
\label{fig:model}
\end{figure}

Motivated by these hot topics on spin-ice systems, we explore
 here another prototype of frustrated 
classical spin system, possessing the {\it garnet structure}, 
which is one of the most popular three-dimensional 
frustrated lattices. The garnet 
lattice is shown in Fig. \ref{fig:model} schematically, which
 is characterized by a specific network of corner-sharing
triangles. In this sense, it should share some essential properties
with the Kagome lattice formed by a two-dimensional
network of corner-sharing
triangles. Also, we expect that 
the system exhibits characteristic properties similar to the 
spin ice as observed in the pyrochlore compounds
\cite{Dy2Ti2O7,Dy2Ti2O7_2,Ho2Ti2O7,Siddharthan,spin_ice,
Melko,Siddharthan2,Ho2Ti2O7_2}.

So far, classical Heisenberg models on the garnet lattice have been 
investigated, and the role of geometrical frustration has been discussed
in connection with the compound $\rm Gd_3Ga_5O_{12}$\cite{GGG1,GGG2,GGG3}.
In this paper, we focus on the Ising spin system, which should 
provide another interesting aspect of frustration in  the garnet 
lattice system. 
We shall indeed uncover that the garnet Ising system  shows
either spin-ice like behavior as the pyrochlore system, or 
magnetic phase transitions depending on the shape of the
interactions. Although experimental study on 
the spin-ice behavior 
on the garnet lattice has not been reported so far, 
we expect that relevant compounds can be synthesized
in the future, adding another interesting example
to the class of spin-ice systems.

This paper is organized as follows. 
In \S \ref{sec:Model}, we introduce the Ising Hamiltonian 
on the garnet lattice, and 
 briefly outline the numerical procedure.
In \S \ref{sec:res}, we first discuss the effects of frustration 
due to the nearest-neighbor 
exchange interaction to explore spin-ice like behavior, and then 
investigate what kind of
magnetic orders are stabilized by the long-range dipolar-type
 interaction.
Brief summary is given in \S \ref{sec:Summary}.

\section{Model and Method}\label{sec:Model}

We investigate a frustrated Ising-spin model on the garnet lattice
 shown in Fig. \ref{fig:model}.
The garnet lattice  composed of corner-sharing triangles possesses
 twenty-four spins in the unit cell.
If one takes into account only the nearest neighbor interaction,
the lattice is divided into two equivalent sublattices,
as shown in Fig. \ref{fig:model} (b).
In some kinds of frustrated spin systems, 
strong easy-axis anisotropy  realizes the Ising spin system
quantized in several distinct directions.
In this paper, we assume strong easy-axis anisotropy 
along the line joining the triangular center,
as realized in spin-ice pyrochlore systems.\cite{Siddharthan}
Define here the notation of scalar Ising spins:
$\sigma_i= 1(-1)$ when a spin points outward (inward) of triangles 
drawn as the gray ones in Fig. \ref{fig:model} (a).
The corresponding unit vector expressing the direction of 
the Ising spin is denoted as ${\mib n}_{i}$.

The Hamiltonian we consider here is
\begin{eqnarray}
{\cal H}&=&\frac{1}{2}\sum_{i,j}\left[J_s\delta_{n.n.}+
J_{ij}\right]\sigma_{i}\sigma_{j}-
\sum_{i}\mu\sigma_{i}{\mib n}_{i}\cdot{\mib H},\nonumber\\
J_{ij}&=&\frac{\mu_0\mu^2}{4\pi}\left[\frac{{\mib n}_i\cdot 
{\mib n}_j}{r_{ij}^3}-3\frac{({\mib n}_i\cdot {\mib r}_{ij})({\mib n}_j
\cdot {\mib r}_{ij})}{r_{ij}^5}\right],\label{1.2}
\end{eqnarray}
where $\mu$ is the magnetic moment, 
$J_s$ is the exchange interaction between nearest neighbor 
spins ($\delta_{n.n.}$ restricts the 
summation only for the nearest neighbors),
$J_{ij}$  is the dipolar interaction between two spins 
separated from each other by the distance $r_{ij}$, 
 and $\mib H$ is the magnetic field.

To investigate magnetic properties of the model,
we make use of Monte Carlo simulations,
where the heat bath method is used to realize the detailed valance.
We perform the simulations for $L\times L\times L$ lattices 
($24L^3$ lattice points) with $L=5,10,15,20$
by imposing periodic boundary conditions. 
To realize the equilibrium state, 
the simulation for the initial relaxation has been made with about
100,000 Monte Carlo steps per spin
to obtain the expectation values for various static quantities 
such as the susceptibility, the specific heat and the entropy.

\section{Results}\label{sec:res}

We first discuss the effects of the nearest-neighbor exchange
interaction $J_s$ to observe how the spin-ice like properties 
show up in the Ising model on the garnet lattice.
We then explore the competition between the nearest-neighbor interaction
$J_s$  and the long-range dipolar interaction $J_{ij}$, and
 determine the phase diagram of the model.

\subsection{Garnet lattice with nearest neighbor interaction}

\begin{figure}[htb]
\begin{center}
\includegraphics[width=7cm]{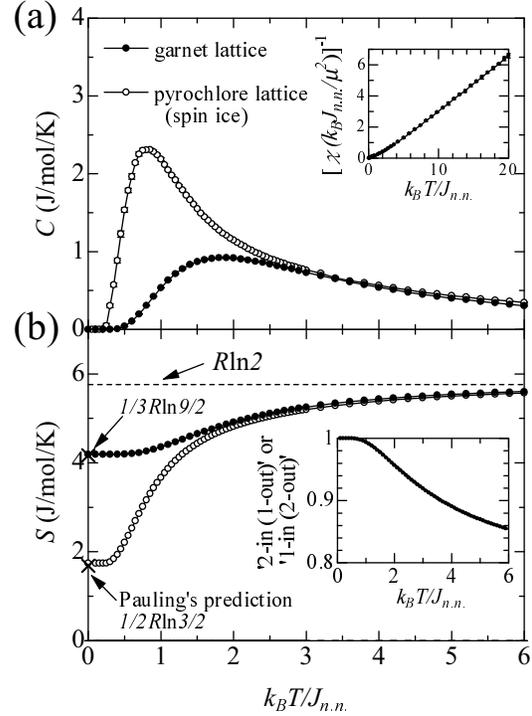}
\end{center}
\caption{(a) Specific heat and (b) entropy of the Ising model 
on the garnet lattice with the nearest-neighbor 
exchange interaction ($J_s>0$)
as a  function of the temperature $T$. 
The results for the pyrochlore lattice (spin ice model) is 
also shown for reference. 
Inset in (a) shows the inverse of the magnetic susceptibility in 
the field direction $[100]$, and inset in (b) shows the probability for
the configurations of two-in (one-out) or one-in (two-out) to
appear. Two crosses are the 
approximate values of the residual entropy
calculated by the method of Pauling:\cite{Pauling} 
$(1/3)R {\rm ln} {9/2}$ for the garnet lattice
and $(1/3)R {\rm ln} {3/2}$ for the pyrochlore lattice
($R=k_B N$).
}
\label{fig:nnCS}
\end{figure}

We start with a garnet system 
with the nearest neighbor exchange interaction $(J_{ij}=0)$.
 Here, we assume 
the exchange coupling, $J_s > 0$, because the model with $J_s < 0$
 without the dipolar interaction leads to a simple ordered phase.
For later convenience, we introduce the effective
coupling, $J_{n.n.}=J_S+J_{D1}$, where $J_{D1}$ is the dipolar 
interaction  between nearest-neighbor spins.
By performing Monte Carlo simulations for the system $(L=10)$, 
we estimate the temperature-dependent specific heat and entropy, 
as shown in Fig. \ref{fig:nnCS}.
At high temperatures, all the spin configurations on each
triangle, i.e. three-in, three-out, two-in (one-out) and one-in (two-out), 
are equally populated. 
As decreasing the temperature, the specific heat features
a Schottky-type  hump  around $k_B T/J_{n.n.}\sim 2$, implying that
a high-temperature  disordered spin phase
gradually changes to the  low-temperature phase 
without any  phase transition. Around the crossover temperature, 
three-in and three-out configurations on each triangle are 
suppressed with the decrease of the temperature. 
This is indeed seen in inset of Fig. \ref{fig:nnCS} (b), where 
  the  two-in (one-out) or one-in (two-out) spin configuration  
gets dominant at low temperatures.  Correspondingly,
 the magnetic susceptibility is enhanced, and diverges 
at absolute zero, as seen in the inset of Fig. \ref{fig:nnCS} (a),
which suggests  the existence of free spins in the ground state.
In fact, Fig. \ref{fig:nnCS} (b) confirms the  finite  residual
entropy  at $T=0$.  Therefore,
 the nearest neighbor Ising model on the garnet lattice 
has the macroscopic ground-state degeneracy 
with large residual entropy. 
These magnetic properties are essentially same as those for
the spin-ice observed in the pyrochlore Ising-spin system.
\cite{Dy2Ti2O7,Siddharthan,spin_ice}
As shown in Fig. \ref{fig:nnCS}, the residual entropy in
 the garnet spin system is larger than in the pyrochlore spin system.
This may reflect the difference in the number of neighbors, which 
is smaller  in the garnet lattice than in the pyrochlore lattice. 

\begin{figure}[htb]
\begin{center}
\includegraphics[width=7cm]{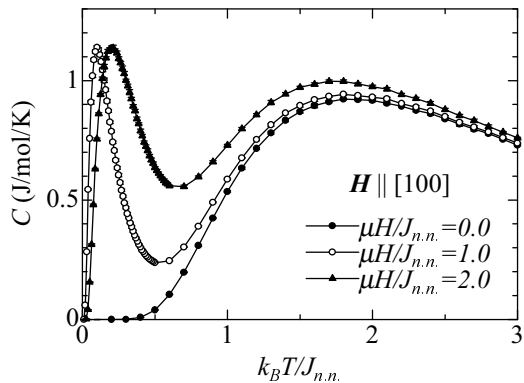}
\end{center}
\caption{
Specific heat as a function of the temperature for 
 various fields applied along the [100] direction.
}
\label{fig:nnCvsH}
\end{figure}

\begin{figure}[htb]
\begin{center}
\includegraphics[width=7cm]{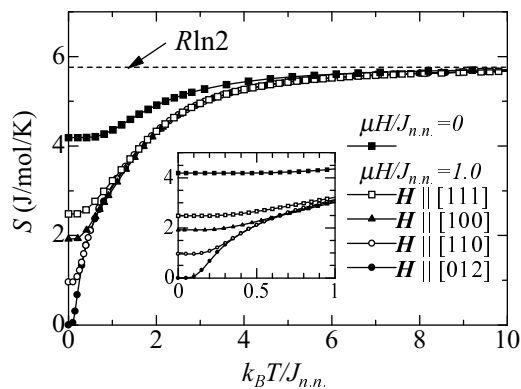}
\end{center}
\caption{Entropy as a function of the 
temperature for the field $H=1.0$.
 The exact value of the residual entropy is $S/R=1/6\ln 6, 
1/3\ln 2, 1/6\ln 2$ and $0$  for the  field
$\mib H \parallel [111], [100], [110]$, and $[012]$, respectively.
These values are obtained exactly by counting the number of
relevant free spins for each case.
}
\label{fig:nnSvsH}
\end{figure}

Shown in  Figs. \ref{fig:nnCvsH} 
and \ref{fig:nnSvsH} are  the data in the presence of 
a magnetic field.
Introducing the field, a new peak structure is developed in the
specific heat at low temperatures, because some 
portion of the ground-state degeneracy is lifted by the 
applied magnetic field. 
Since  there are several distinct Ising axes in 
the garnet lattice,  the residual entropy takes different values
depending on the direction of the 
applied  field, as shown in Fig.\ref{fig:nnSvsH}.

Most of the above properties are similar to those observed for  
 the pyrochlore spin-ice system 
\cite{Dy2Ti2O7,spin_ice,kagome1,kagome2,kagome3}
except that
in the present model the residual entropy is large, and an
applied field produces a variety of ground states, which 
still have macroscopic degeneracy.

\subsection{Effects of long-range  interaction}
Having seen that the model 
with the nearest neighbor interaction exhibits 
 spin-ice like properties, we now ask 
 how stable the  ground state is when the
additional interaction among spins is introduced.
In fact, according to recent reports for some pyrochlore 
compounds such as $\rm Dy_2Ti_2O_7$ and
$\rm Ho_2Ti_2O_7$, the dipolar interaction may 
be rather important so as to enhance the tendency to
stabilize a magnetically ordered ground state.\cite{Melko,Ewald2}

In the following, we consider the long-range 
dipolar interaction  to observe the stability of
 the spin-ice like state  against possible ordered states. 
To analyze the effect of the dipolar interaction systematically, 
we follow the way of ref.\cite{Siddharthan} and introduce 
the cutoff $n_c$  to control  the effective
range of the interaction, where the interactions up to $n_c$th
neighbor spins are taken into account.
Although the cutoff  has little effect 
on the phase transition in the pyrochlore system,\cite{Siddharthan}
the present garnet lattice system may be more sensitive 
to the cutoff  since the degeneracy of the ground state 
is much larger than the pyrochlore system.

We study the system with various choices of the cutoff
up to $n_c=20$.  The case with $n_c=\infty$ is treated 
by exploiting a different technique, as will be mentioned 
later in this section.
Since the typical behavior found
for the phase transitions is 
classified into three categories, 
we show the results for three representative examples by
setting the cutoff as
$n_c=4$, 5 and 14, which respectively 
correspond to the real cutoff distance 
$l_c/a=0.56$, 0.59 and  1.0 in unit of the lattice constant $a$.
For a while, we restrict ourselves to the case with $J_s=0$, and 
 discuss the effect of $J_s$ later in this section.

Before starting the discussions on the phase transitions, we mention
here that the models having the interaction with $n_c=1, 2,3$ show
the spin-ice like behavior analogous to the 
nearest-neighbor model.


\subsubsection{second-order transition {\rm ($n_c=4, 9, 10,11$)}}

\begin{figure}[htb]
\begin{center}
\includegraphics[width=7cm]{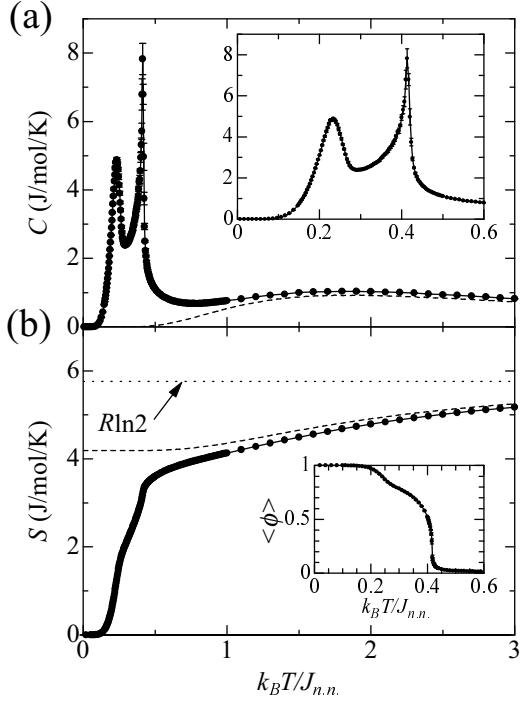}
\end{center}
\caption{Specific heat and entropy as a function of the temperature
for the model having  the interactions up to fourth-neighbors.
The specific heat for the nearest-neighbor model is shown as the 
broken line for reference.
Inset in (a) magnifies the specific heat in the low
temperature region, and inset in (b) 
shows the temperature dependence of the 
 order parameter $\phi$.
} 
\label{fig:4th}
\end{figure}

We start with the
results for the model including interaction up to
 fourth-neighbors ($n_c=4$)  shown in Fig. \ref{fig:4th}.
It is found that two peaks appear at low temperatures besides
a much broader hump at higher temperatures 
around $(k_BT/J_{n.n.}\sim 2)$.
The higher-temperature hump is essentially the same as
 that for the model with the nearest neighbor interaction.
Therefore, we can see that two peaks at low temperatures are featured by 
lifting the ground-state degeneracy due to the newly introduced interactions.
These two peaks are different in character as seen in the inset of
Fig. \ref{fig:4th} (a). 

\begin{figure}[htb]
\begin{center}
\includegraphics[width=7cm]{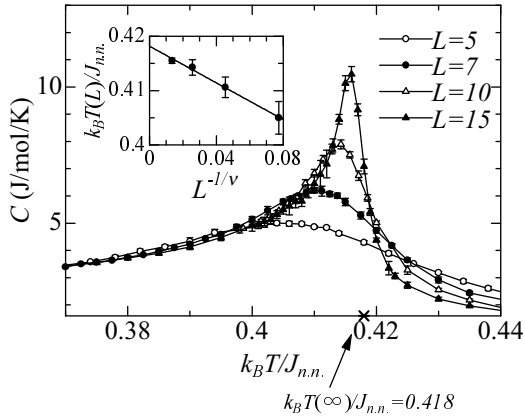}
\end{center}
\caption{Specific heat for different choices of  the system size. 
Inset shows the scaling plot of the peak position, which should 
give the critical temperature in the thermodynamic limit.
}
\label{fig:size}
\end{figure}

The higher-temperature peak seems to show divergent behavior, implying 
a second-order phase transition.
We indeed confirm this by examining the system-size dependence
of the  specific heat systematically,
as shown in Fig. \ref{fig:size}.
It is found that the increase of the system size enhances the peak
and slightly changes  the location of the peak.
As shown in the inset of Fig. \ref{fig:size}, 
the peak position for each system size satisfies
the scaling plot with the Ising critical exponent $\nu=0.629$.
Therefore, we conclude that
this phase transition belongs to the universality class 
of the three-dimensional Ising model,\cite{critical}
and the critical temperature thus 
determined is  $k_BT_c/J_{n.n}=0.418$. 

\begin{figure}[htb]
\begin{center}
\includegraphics[width=7cm]{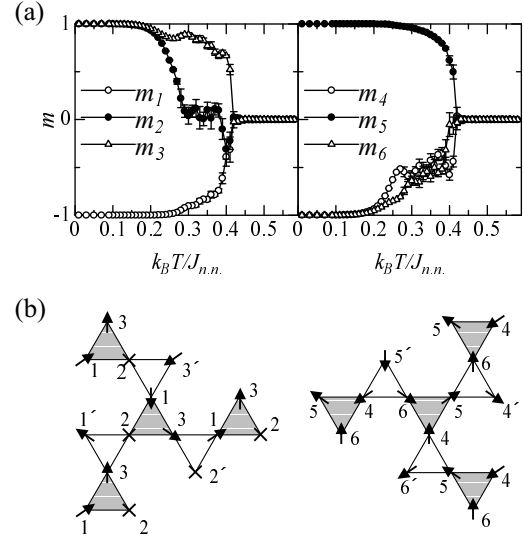}
\end{center}
\caption{(a) Spontaneous magnetization for each sublattice $m_k$ as 
a function of the temperature $T$. It is seen that just below the critical
temperature, the magnetization
is not developed for one of six sublattices (lattice 2 in this
example). 
(b) Spin-configuration  pattern for 
the partially disordered phase, where crosses on the lattice 2 mean
 that spins are still free there.
}
\label{fig:order+}
\end{figure}

We now wish to ask what kind of order is realized at this 
phase transition.  We find that the system has 
the finite order parameter $\phi$, which is defined as,
\begin{eqnarray}
\langle\phi\rangle&=&\frac{1}{4}\langle(m_1-m_2)^2
+(m_2-m_3)^2+(m_3-m_1)^2\nonumber\\
&&+(m_4-m_5)^2+(m_5-m_6)^2+(m_6-m_4)^2\rangle^{\frac{1}{2}},\nonumber\\
\\
m_k&=&\frac{6}{N}\sum_{\substack{i \\ over \, kth \\sublattice}}
\sigma_i,
\end{eqnarray}
where $m_k$ is the magnetization of the $k$th sublattice
($k=1 \sim  6$).
In the inset of Fig. \ref{fig:4th}(b), we show how this
order parameter is developed.
In order to further clarify the detailed structure of the 
ordered state,
 we show the spontaneous magnetization for each 
sublattice in Fig. \ref{fig:order+}.
The analysis of the magnetization reveals that below the 
critical temperature
the symmetry breaks only partially, namely, there appears
a finite magnetization in five kinds of the sublattices, but
no magnetization in the remaining one sublattice.
The corresponding spin arrangement 
is drawn in Fig. \ref{fig:order+} (b) schematically.
Therefore, we can say that a {\it partially disordered phase}
is realized below the critical temperature $(k_BT/J_{n.n.}< 0.4)$.
This state is continuously connected to 
the completely ordered phase realized at zero temperature
without any phase transition.  This crossover naturally explains
why a Schottky-like structure is developed in 
the specific heat at low temperatures, as observed in Fig. \ref{fig:4th}.
We have confirmed that this type of the second-order transition 
occurs for the model with $n_c=4, 9-11$.



\subsubsection{double second-order transitions {\rm  ($n_c=5, 6, 7$)} }

We next discuss another prototype of phase transition, 
by taking  the $n_c=5$ case as an example.  The analysis
can be done in a similar way mentioned above, so that we 
briefly summarize the relevant points below.

\begin{figure}[htb]
\begin{center}
\includegraphics[width=6.6cm]{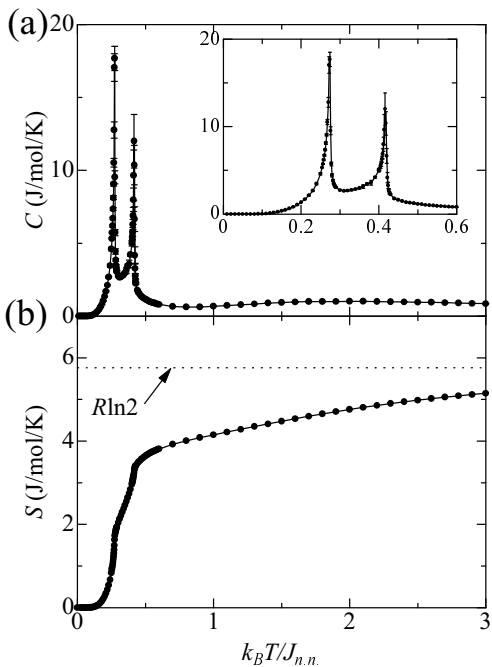}
\end{center}
\caption{(a) Specific heat and (b) entropy for the model having
the fifth-neighbor interactions ($n_c=5$).
}
\label{fig:5thCS+}
\end{figure}

\begin{figure}[htb]
\begin{center}
\includegraphics[width=7cm]{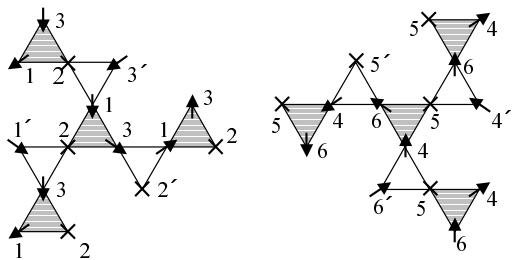}
\end{center}
\caption{Spin-configuration pattern of  the intermediate 
ordered phase  for the model having the interaction with $n_c=5$.
There are specific correlations between left and right panels
(see text).
}
\label{fig:5thpd+}
\end{figure}

In Fig. \ref{fig:5thCS+}, we show the results for 
the specific heat and the entropy.
As is the case for the model with  $n_c=4$, 
the temperature-dependent 
specific heat has a wide hump at high temperatures, which is 
supplemented by two peaks at low temperatures.  In this case, however, 
both of the two peaks show divergent behavior, implying 
that the second-order phase transition takes place twice.
We would naively expect that the crossover behavior characterized 
by the lower peak in the $n_c=4$ case  now changes its character to 
the second-order phase transition.  However, it turns out that this simple 
picture is not correct. By calculating the spontaneous magnetization,
we find that in the intermediate phase ($0.27 <k_BT/J_{n.n.}< 0.42$), 
a partially disordered phase is realized  with  spin correlations
between different sublattices,
shown schematically in Fig. \ref{fig:5thpd+}.
Namely, there are specific correlations between the configulations
 in the left and right panels:
for example, if the lattice sites labeled by 2 in the left
 are disordered, so that the corresponding lattice sites labeled by 5 
in the right panel should be disordered.  As another 
example, if we assume the lattice
sites labeled  by 1 to be disordered, then the lattice sites labeled
by  4
should be disordered. As decreasing the temperature, this partially
disordered phase is driven to the completely ordered phase
via another second-order phase transition.  We have checked that
this type of phase transitions are realized  for the 
models with $n_c=5, 6$ and 7.


\subsubsection{first-order transition {\rm ($n_c=8, \,\, n_c \ge 12$)} }

We finally discuss the case exhibiting the first-order phase
transition, by taking the case of $n_c=14$,
where the interaction is taken into account 
up to the fourteenth neighbors.

\begin{figure}[htb]
\begin{center}
\includegraphics[width=7cm]{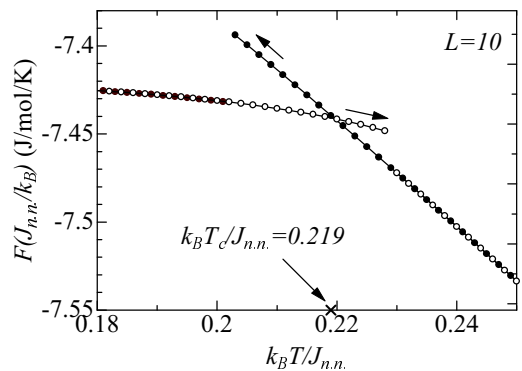}
\end{center}
\caption{Free energy $F$ for the Ising model with $n_c=14$,
which clearly shows a cusp singularity.
Solid (open) circles represent the results obtained 
when the temperature is decreased (increased).
}
\label{fig:free}
\end{figure}

To see the nature of the first-order phase transition clearly, we
 first look at the temperature-dependent free energy 
shown in Fig. \ref{fig:free}.
As decreasing (increasing) the temperature from higher 
(lower) temperatures, the free energy features distinct
curves, as is typical for first-order transitions.
We note here that
in actual Monte Carlo calculations for a finite-size system, 
there appears hysterisis, which is
 accompanied by a sudden jump of the free energy at a 
certain temperature. Nevertheless, 
we can determine the  critical temperature, 
$k_B T/J_{n.n.}\sim 0.219$, correctly from the crossing point in
 the free energy curves.

\begin{figure}[htb]
\begin{center}
\includegraphics[width=6.6cm]{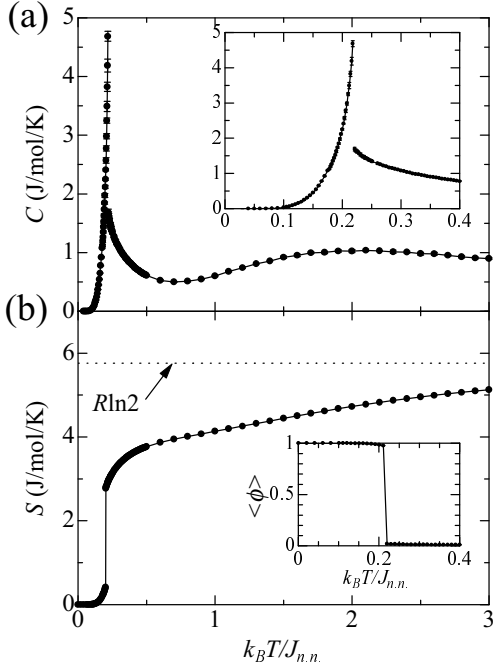}
\end{center}
\caption{
(a) Specific heat and (b) entropy as a function of the temperature
for the system with $n_c=14$.
Inset in (b) shows the temperature dependence of the 
order parameter $<\phi>$.
}
\label{fig:hys}
\end{figure}

The specific heat and entropy calculated are shown in Fig. \ref{fig:hys}.
In the high temperature region, a broad hump appears again 
in the specific heat around $k_BT/J_{n.n.}\sim 2$, similarly to the previous 
two cases, reflecting the crossover from the free spin state to
 the spin ice state. In the present case, the first-order transition
 gives rise to the discontinuity  at the critical point
in the specific heat, the entropy and  the order parameter, as
seen  in Fig. \ref{fig:hys}.
A remarkable point is that the ordered state below the critical
 temperature
has  the same order as for the $n_c=4$ case but not for the
$n_c=5$ case.  This is a nontrivial consequence due to
the long-range interaction on the frustrated garnet lattice.

We have confirmed that the above type of the first-order phase transition
occurs for the model with $n_c =8$ and $n_c\ge 12$.
In this connection,
 we make a comment on the system with the original
 dipolar interaction ($n_c=\infty$). Since our technique 
employed here is not appropriate to treat the $n_c=\infty$ case
directly, we have instead used the Ewald
 method,\cite{Ewald1,Melko,Ewald2} which can overcome the above
difficulty. We have found that the first-order transition 
similar to Fig. \ref{fig:hys} indeed occurs even in the $n_c=\infty$
case.

\subsection{Phase diagram}

\begin{figure}[htb]
\begin{center}
\includegraphics[width=7cm]{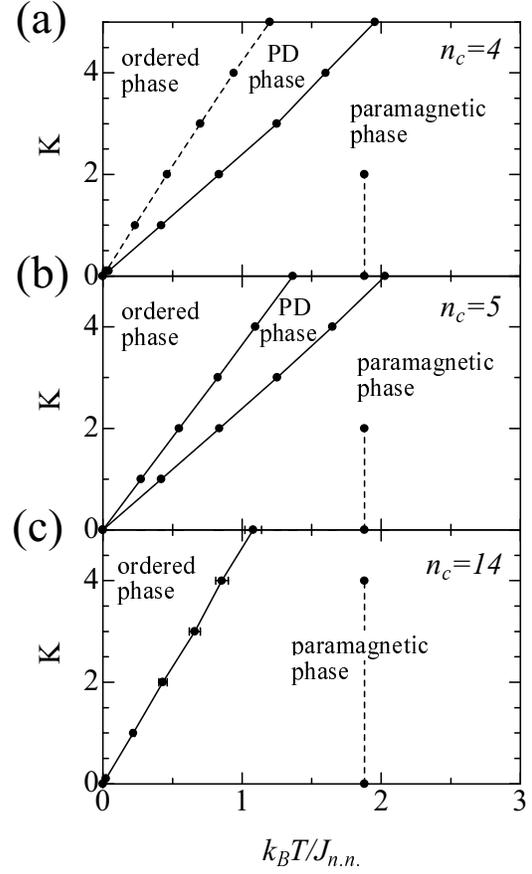}
\end{center}
\caption{Phase diagrams for the Ising spin system on the garnet lattice,
where the long-range dipolar interaction is taken into account 
up to  $n_c=4$, 5 and 14  neighbors.
Solid lines represents the phase boundary
 between the magnetically ordered phase 
and the paramagnetic phase (second-order transition
 for (a) and (b), and first-order transition for (c)).
Dashed lines represent the crossover around which 
the specific heat exhibits its maximum value.
}
\label{fig:diagram}
\end{figure}

We repeat similar calculations by changing the ratio of the 
exchange interaction and  the long-range interaction
to discuss the competition between the spin-ice like state and 
the magnetically ordered states. Here, we make use of the parameter
$K=J_{D1}/J_{n.n}$  with $J_{n.n.}=J_S+J_{D1}$  defined before.
We show the phase diagrams obtained
for the three prototype cases
with $n_c=4$ and 5, 14 in Fig. \ref{fig:diagram}.
As mentioned above, 
when $K=0$ $(J_{ij}=0)$, the system is reduced to the Ising model 
with nearest neighbor interactions.
It is seen that the introduction of the long-range interaction
has little effect on the spin-ice like behavior at higher temperatures.
Namely, the temperature characterizing the crossover is 
little changed by the long-range interaction. This is because 
spin-ice like behavior  is dominated mainly  by 
local spin fluctuations. 
On the other hand, the long-range interaction enhances 
long-range correlations among spins in the same sublattice, 
inducing the phase transition to several distinct 
 magnetically ordered ground states 
at low temperatures. In particular, we have obtained
partially disordered phases, which are inherent in frustrated
Ising systems\cite{PD1,PD2,PD3,PD4}.

In comparison with the  pyrochlore system with the long-range
dipolar interaction\cite{Siddharthan}, the 
present system shows rather complicated
phase diagrams, depending on the cutoff distance.
This may be related to the fact that the garnet
system has much larger residual entropy than that for the pyrochlore
system.

In realistic Ising garnet systems, it
 may be expected that either the spin-ice 
behavior or the 
first-order phase transition would occur.

\section{Summary}\label{sec:Summary}

We have investigated the classical Ising model 
on the garnet lattice
by means of Monte Carlo simulations with the heat bath algorithm.
It has been clarified that the Ising model with nearest neighbor 
 interaction exhibits the spin-ice like behavior without any 
phase transitions,
resulting in the large residual entropy at zero temperature.
It has been found that a part of the degeneracy in the ground state 
is lifted upon introducing a magnetic field.
We have also discussed the effect of the dipolar-type interaction
to clarify how the spin-ice like state competes with 
the magnetically ordered states.

We have seen that when  the long-range dipolar interaction exists
the first-order transition takes place even though the magnitude 
of the dipolar interaction is small.  However, the transition temperature 
should be very low in such cases, making
the spin-ice like behavior  dominant in the wide temperature 
range.

Although  the spin ice behavior  has not been observed 
experimentally in the existing garnet compounds, we expect that 
possible candidates can be synthesized in the future, 
which may further stimulate theoretical and experimental 
investigations in this class of spin-ice systems.


\section*{Acknowledgement}
This work was partly supported by a Grant-in-Aid from the Ministry 
of Education, Science, Sports and Culture of Japan.

\end{document}